\RequirePackage[2020-02-02]{latexrelease}
\documentclass[
  twocolumn,
  showpacs,
  preprintnumbers,
  amsmath,
  amssymb,
  aps,
  prc,
  floatfix,
  letterpaper,
  superscriptaddress,nofootinbib
]{revtex4-1}
\makeatletter
\def\pdfstartlink@attr{}
\makeatother
\usepackage{tabularx}
\usepackage{graphicx}
\usepackage{dcolumn}     
\usepackage{bm}          
\usepackage[hidelinks]{hyperref}
\usepackage{parskip}
\def\today{\ifcase\month\or
  January\or February\or March\or April\or May\or June\or
  July\or August\or September\or October\or November\or December\fi
  \space\number\day, \number\year}

\newcommand\NSU{Norfolk State University, Norfolk, Virginia 23504, USA}
\newcommand\UNH{University of New Hampshire, Durham, New Hampshire 03824, USA}
\newcommand\UVA{University of Virginia, Charlottesville, Virginia 22903, USA}
\newcommand\WM{The College of William and Mary, Williamsburg, Virginia 23187, USA}
\newcommand\CAT{Catholic University of America, Washington, DC 20064, USA}
\newcommand\Jlab{Thomas Jefferson National Accelerator Facility,
  Newport News, Virginia 23606, USA }
\newcommand\Rutgers{  Rutgers University, New Brunswick, NJ 08901, USA}
\newcommand\MIT{Massachusetts Institute of Technology, Cambridge, MA 02139, USA}
\newcommand\XLA{Xavier University of Louisiana, New Orleans, LA 70125, USA}
\newcommand\Seoul{Seoul National University, Seoul 151-742, South Korea}
\newcommand\INFN{Istituto Nazionale di Fisica Nucleare, Sezione di Roma Tor Vergata, I-00173  Rome, Italy}

\newcommand\Catania{Universita di Catania, 95124 Catania, Italy}
\newcommand\USC{University of South Carolina, Columbia, SC 29208, USA}
\newcommand\Temple{Temple University, Philadelphia, Pennsylvania 19122, USA}
\newcommand\Argonne{Argonne National Laboratory, Argonne, Illinois 60439, USA}
\newcommand\CalLA{California State University, Los Angeles, Los Angeles, California 90032, USA}
\newcommand\Orsay{Institut de Physique Nucleaire, 91400 Orsay, France}
\newcommand\FIU{Florida International University, Miami, Florida 33199, USA}
\newcommand\USTC{University of Science and Technology, Hefei 230000, China}
\newcommand\CAS{Chinese Academy of Sciences, Beijing 100045, China}
\newcommand\HamptonU{Hampton University, Hampton, Virginia 23668, USA}
\newcommand\ODU{Old Dominion University, Norfolk, Virginia 23529, USA}
\newcommand\CNU{Christopher Newport University, Newport News, VA 23606, USA}
\newcommand\TelAviv{Tel Aviv University, Tel Aviv, 69978 Israel}
\newcommand\Yerevan{Yerevan Physics Institute, Yerevan, Armenia 0036}
\newcommand\Stefan{Jo\v{z}ef Stefan Institute, Ljubljana, Slovenia}
\newcommand\Hebrew{Hebrew University, Jerusalem 9190401, Israel}
\newcommand\Duke{Duke University, Durham, NC 27708, USA}
\newcommand\Glasgow{Glasgow University, Glasgow G12 8QQ, Scotland}
\newcommand\ORNL{Oak Ridge National Laboratory Oak Ridge, TN 37830, USA}
\newcommand\Kharkov{Kharkov Institute of Physics and Technology, Kharkov 61108, Ukraine}
\newcommand\Regina{University of Regina, Regina, Canada}
\newcommand\Mississippi{Mississippi State University, Mississippi State, MS
39762, USA}
\newcommand\Berkeley{Lawrence Berkeley National Laboratory, Berkeley, CA 94720, USA}
\newcommand\Ljubljana{Faculty of Mathematics and Physics, University of Ljubljana, Ljubljana, Slovenia}

\renewcommand{\thefootnote}{\fnsymbol{footnote}}

\begin{document}
\setcounter{footnote}{1}
\footnotetext{Corresponding author: slifer@jlab.org}
\title{The Proton Spin Structure Function \texorpdfstring{$g_2$}{g2} and Generalized Polarizabilities in the Strong QCD Regime}
\author{D. Ruth} \affiliation{\UNH}
\author{R.~Zielinski} \affiliation{\UNH}
\author{C.~Gu} \affiliation{\UVA}
\author{M.~Allada~(Cummings)} \affiliation{\WM}
\author{T.~Badman} \affiliation{\UNH}
\author{M.~Huang} \affiliation{\Duke}
\author{J.~Liu} \affiliation{\UVA}
\author{P.~Zhu} \affiliation{\USTC}
\author{K.~Allada} \affiliation{\MIT}
\author{J.~Zhang} \affiliation{\Jlab}
\author{A.~Camsonne} \affiliation{\Jlab}
\author{J.~P.~Chen} \affiliation{\Jlab}
\author{K.~Slifer$^\dagger$} \affiliation{\UNH}
\author{K.~Aniol} \affiliation{\CalLA}
\author{J.~Annand} \affiliation{\Glasgow}
\author{J.~Arrington} \affiliation{\Argonne} \affiliation{\Berkeley}
\author{T.~Averett} \affiliation{\WM}
\author{H.~Baghdasaryan} \affiliation{\UVA}
\author{V.~Bellini} \affiliation{\Catania}
\author{W.~Boeglin} \affiliation{\FIU}
\author{J.~Brock} \affiliation{\Jlab}
\author{C.~Carlin} \affiliation{\Jlab}
\author{C.~Chen}\affiliation{\HamptonU}
\author{E.~Cisbani} \affiliation{\INFN}
\author{D.~Crabb} \affiliation{\UVA}
\author{A.~Daniel} \affiliation{\UVA}
\author{D.~Day} \affiliation{\UVA}
\author{R.~Duve} \affiliation{\UVA}
\author{L.~El~Fassi} \affiliation{\Rutgers} \affiliation{\Mississippi}
\author{M.~Friedman} \affiliation{\Hebrew}
\author{E.~Fuchey} \affiliation{\Temple}
\author{H.~Gao} \affiliation{\Duke}
\author{R.~Gilman} \affiliation{\Rutgers}
\author{S.~Glamazdin} \affiliation{\Kharkov}
\author{P.~Gueye} \affiliation{\HamptonU}
\author{M.~Hafez} \affiliation{\ODU}
\author{Y.~Han} \affiliation{\HamptonU}
\author{O.~Hansen} \affiliation{\Jlab}
\author{M.~Hashemi Shabestari} \affiliation{\UVA}
\author{O.~Hen} \affiliation{\MIT}
\author{D.~Higinbotham} \affiliation{\Jlab}
\author{T.~Horn} \affiliation{\CAT}
\author{S.~Iqbal} \affiliation{\CalLA}
\author{E.~Jensen} \affiliation{\CNU}
\author{H.~Kang} \affiliation{\Seoul}
\author{C.~D.~Keith} \affiliation{\Jlab}
\author{A.~Kelleher} \affiliation{\MIT}
\author{D.~Keller} \affiliation{\UVA}
\author{H.~Khanal} \affiliation{\FIU}
\author{I.~Korover} \affiliation{\TelAviv}
\author{G.~Kumbartzki} \affiliation{\Rutgers}
\author{W.~Li} \affiliation{\Regina}
\author{J.~Lichtenstadt} \affiliation{\TelAviv}
\author{R.~Lindgren} \affiliation{\UVA}
\author{E.~Long} \affiliation{\UNH}
\author{S.~Malace} \affiliation{\USC}
\author{P.~Markowitz} \affiliation{\FIU}
\author{J.~Maxwell} \affiliation{\UNH} \affiliation{\Jlab}
\author{D.~M.~Meekins} \affiliation{\Jlab}
\author{Z.~E.~Meziani} \affiliation{\Temple}
\author{C.~McLean} \affiliation{\WM}
\author{R.~Michaels} \affiliation{\Jlab}
\author{M.~Mihovilovi\v{c}} \affiliation{\Ljubljana} \affiliation{\Stefan}
\author{N.~Muangma} \affiliation{\MIT}
\author{C.~Munoz Camacho} \affiliation{\Orsay}
\author{J.~Musson} \affiliation{\Jlab}
\author{K.~Myers} \affiliation{\Rutgers}
\author{Y.~Oh} \affiliation{\Seoul}
\author{M.~Pannunzio Carmignotto} \affiliation{\CAT}
\author{C.~Perdrisat} \affiliation{\WM}
\author{S.~Phillips} \affiliation{\UNH}
\author{E.~Piasetzky} \affiliation{\TelAviv}
\author{J.~Pierce} \affiliation{\Jlab} \affiliation{\ORNL}
\author{V.~Punjabi}\affiliation{\NSU}
\author{Y.~Qiang} \affiliation{\Jlab}
\author{P.~E.~Reimer}\affiliation{\Argonne}
\author{Y.~Roblin} \affiliation{\Jlab}
\author{G.~Ron} \affiliation{\Hebrew}
\author{O.~Rondon} \affiliation{\UVA}
\author{G.~Russo} \affiliation{\Catania}
\author{K.~Saenboonruang} \affiliation{\UVA}
\author{B.~Sawatzky} \affiliation{\Jlab}
\author{A.~Shahinyan} \affiliation{\Yerevan}
\author{R.~Shneor} \affiliation{\TelAviv}
\author{S.~\v{S}irca} \affiliation{\Ljubljana} \affiliation{\Stefan}
\author{J.~Sjoegren} \affiliation{\Glasgow}
\author{P.~Solvignon-Slifer} \affiliation{\UNH}
\author{N.~Sparveris} \affiliation{\Temple}
\author{V.~Sulkosky} \affiliation{\MIT}
\author{F.~Wesselmann} \affiliation{\XLA}
\author{W.~Yan} \affiliation{\USTC}
\author{H.~Yang} \affiliation{\CAS}
\author{H.~Yao} \affiliation{\WM}
\author{Z.~Ye} \affiliation{\UVA}
\author{M.~Yurov} \affiliation{\UVA}
\author{Y.~Zhang} \affiliation{\Rutgers}
\author{Y.~X.~Zhao} \affiliation{\USTC}
\author{X.~Zheng} \affiliation{\UVA}

\collaboration{The Jefferson Lab Hall A g2p Collaboration}
\noaffiliation

\date{\today}
             
\begin{abstract}
The strong interaction is not  well understood at low energy, or for interactions with low momentum transfer $Q^2$, but one of the clearest insights we have comes from Chiral Perturbation Theory ($\chi$PT). This effective treatment gives testable predictions for the nucleonic generalized  polarizabilities --- fundamental quantities describing the nucleon's response to an external field. 
We have measured the proton's generalized spin polarizabilities  in the region where  $\chi$PT  is expected to be valid. Our results include the first ever data for the transverse-longitudinal spin polarizability $\delta_{LT}$, and also extend the coverage of the polarizability $\overline{d_2}$ to very low $Q^2$ for the first time. These results were extracted from moments of the structure function $g_2$, a quantity which characterizes the internal spin structure of the proton. Our experiment ran at Jefferson Lab  using a polarized electron beam and a polarized solid ammonia (NH$_3$) target. 
The  $\delta_{LT}$ polarizability  has remained a challenging quantity for $\chi$PT  to reproduce, despite its reduced sensitivity to higher resonance contributions; recent competing calculations still disagree with each other and also diverge from the measured neutron data at very low $Q^2$. Our proton results provide discriminating power between existing calculations, and will help provide a better understanding of this 
strong QCD regime.

\end{abstract}

\maketitle
The proton accounts for the vast majority of the ordinary matter in the universe, but how its fundamental properties such as mass and spin arise from the interactions of its  constituents remains an open question. The proton's sub-structure is well understood in the high energy, short distance region of asymptotic freedom~\cite{Feynman:new}. Conversely, in the low-energy regime where the QCD coupling becomes truly strong, quark-quark and quark-gluon correlations invalidate the simple picture of the parton model.
 Unresolved questions about the structure of the proton can be probed with spin polarized electron-proton scattering, but this region carries a number of experimental challenges that make data scarce. These challenges are especially difficult at low energy for transversely polarized protons. 
 
 It's clear that the nucleons are not yet well understood in this regime. For example, neutron data~\cite{E94010} revealed a large discrepancy between moments of the spin structure functions and early calculations of chiral perturbation theory ($\chi$PT).  This discrepancy became known as the ``$\delta_{LT}$ Puzzle'' and stimulated a rigorous new generation of theoretical efforts and low $Q^2$ experiments such as the present  measurement. Though some light has been shed on the initial disagreement, the newest neutron data~\cite{saGDH} still show a large deviation from $\chi$PT, making it very important to examine if similar issues exist for the proton. Several recently published experiments~\cite{EG4_final,saGDH} have probed the low and medium energy regions of the spin structure function $g_1^p$, and the medium energy region of the spin structure function $g_2^p$. Our experiment expands on those results with low-energy measurements of $g_2^p$ and its associated moments.

 Sum rules  and moments of the nucleon spin structure functions (SSF) allow for a direct comparison between experiment and theory. In recent years, the Bjorken sum rule~\cite{Bjorken:1968dy} at large $Q^2$, and the Gerasimov-Drell-Hearn (GDH) sum rule~\cite{GDH} at $Q^2$ = 0, have been extensively investigated~\cite{COMPASS,Deur:2004ti,PhotonPol,bjorken_deur}. Less well studied for the proton is another class of sum rules addressing the spin polarizabilities~\cite{Drechsel2}. Spin polarizabilities are fundamental properties of the nucleons, making their measurement and comparison to theory of great interest. Polarizabilities, in general, describe the proton's response to an external field. The electric and magnetic polarizabilities are relatively well measured for the proton, but less well understood are the spin and color polarizabilities. Spin polarizabilities describe a spin-dependent response of the nucleon to an electromagnetic field, while color polarizabilities contain information on how the nucleon's spin affects the color electric and magnetic fields on average~\cite{Meziani:2004ne,Gold}. Of particular interest is the spin polarizability $\delta_{LT}$, and the $\overline{d_2}$ higher moment which is related to the color polarizabilities at large $Q^2$.
 
 The sum rules used to obtain these moments are based on unsubtracted dispersion relations, which use the optical theorem to relate moments of the spin structure functions to real or virtual Compton amplitudes~\cite{Chen:2010td}.
 The doubly-virtual Compton scattering dispersion relations are used to form a low-energy expansion of the spin-flip Compton amplitudes $f_{TT}$ and $f_{LT}$ \cite{Drechsel}, giving rise to a number of SSF moments. The next-to-leading order term of the $f_{LT}$ low energy expansion contains the generalized longitudinal-transverse (LT) spin polarizability:
\begin{align}
\label{DLTEQ}
\delta_{LT} (Q^2) = \frac{16\alpha M^2}{Q^6}\int_{0}^{x_0} x^2{\bigg (} g_1(x,Q^2) + g_2(x,Q^2){\bigg )} dx
\end{align}

Here, $Q^2$ is the four-momentum transfer, $\alpha$ is the fine structure constant,  M represents the proton mass, and $x_0$ represents the Bjorken  $\mathnormal{x}$ associated with the pion production threshold at an invariant mass (defined as $W^2 = M^2 +2M\nu -Q^2$) of W=1073.2 MeV.  This limit ensures that the elastic response is excluded from the integral as is required for a pure polarizability~\cite{Gold}.

A measurement of the generalized LT polarizability is considered a benchmark test of $\chi$PT because it is a fundamental nucleon observable, and was initially expected to be insensitive to contributions from virtual $\pi$-$\Delta$ intermediate states~\cite{Bernard3,Kao}. The actual contribution of these states has ultimately proved to be more complicated. The $\chi$PT predictions for $\delta_{LT}$ in LO and NLO are in principle parameter-free predictions, the accuracy of which is determined only by the convergence properties of the chiral expansion. Precise comparisons of this quantity between data and theory are therefore extremely valuable to test $\chi$PT and other low-energy theories.

Also of interest is the $\overline{d_2}$ spin polarizability, 
a higher moment which is identified at high $Q^2$ with the twist-3 matrix element $d_2$~\cite{Meziani:2004ne}. We can access this moment through a sum rule:
\begin{align}
\label{D2EQ}
\overline{d_2} (Q^2) = \int_{0}^{x_0} x^2{\bigg (} 2 g_1(x,Q^2) + 3 g_2(x,Q^2){\bigg )} dx
\end{align}
At high $Q^2$ this moment describes the ``color Lorentz force'' and gives us information on the color polarizabilities discussed above. Consequently, $\overline{d_2}$ helps to describe how the color electric and magnetic fields interact with the nucleon spin~\cite{Meziani:2004ne}. In this regime, $\overline{d_2}$ becomes a twist-3 quantity and thus quantifies the quark-gluon correlations of the nucleon~\cite{Burkardtg2d2}. At low $Q^2$, the partonic description of $\overline{d_2}$ fails, but it remains a pure polarizability which informs the hadronic behavior of the nucleon\cite{Gold,Kaod2}. The moment vanishes for $Q^2$ = 0 and $Q^2$ = $\infty$, but must also transition smoothly through these very different regimes, with phenomenological models suggesting a maximum in the transition region around 1 GeV$^2$~\cite{Osipenko:2005nx}. 
A measurement of $\overline{d_2}$ over a broad region will help to understand the transition between the partonic and hadronic descriptions of the nucleon~\cite{Kaod2}, but data for this observable has proven difficult to obtain. 



In this article, we present the measurement of the proton spin structure function $g_2$ for a range of $Q^2$ 
from 0.02 to 0.13 GeV$^2$. 
We extract the polarizabilities $\delta_{LT}$ 
and $\overline{d_{2}}$ 
 and compare our results 
to the leading predictions of $\chi$PT.

The E08-027 (g2p) experiment~\cite{Zielinski:2017gwp} was performed in Hall A at the Thomas Jefferson National Accelerator Facility (JLab). We performed an inclusive measurement at forward angles of the proton spin-dependent cross sections. A longitudinally polarized electron beam was scattered from a longitudinally or transversely polarized solid  NH$_3$ target. Data were collected at three different beam energies and two different target field strengths. The transverse field of the transversely polarized NH$_3$ target influenced the scattered electron trajectory sufficiently to obtain two separate $Q^2$ values at a single beam energy for different target magnetic field strengths. In total, the results cover five transverse field kinematic settings and one longitudinal field setting, a table of which is found in the supplemental materials. The measurements covered values of the invariant mass from the nuclear elastic peak through the nucleon resonance region, but only the results above the pion production threshold ($W$= 1073.2 MeV) are discussed in this letter.
 \begin{figure}[t!]
\includegraphics[angle=0,width=0.48\textwidth]{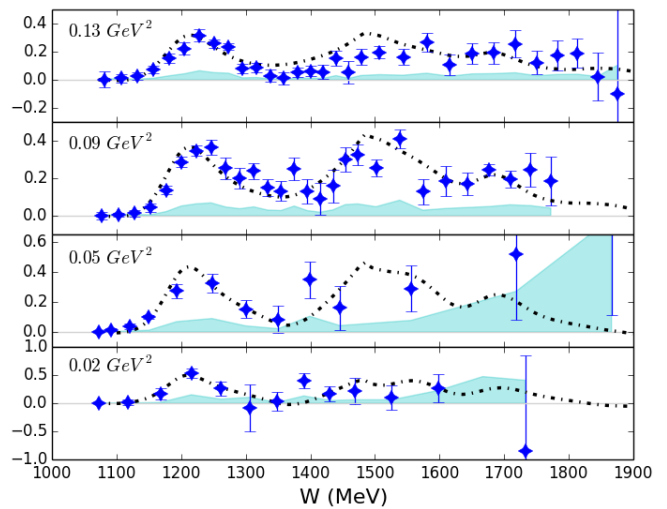}
\caption{\label{fig:SSF} The proton spin structure function $g_2$ as a function of invariant mass $W$ at the constant $Q^2$ designated in the upper left of each panel. The error bars are statistical and the shaded region represents the  the systematic uncertainty. The black dashed line represents the phenomenological Hall B model~\cite{HallB}.
}
\end{figure}

We extracted the spin structure functions from a calculation of the polarized cross section differences $\Delta \sigma_{\parallel} = \frac{d^2\sigma}{d\Omega dE'} (\downarrow \Uparrow - \uparrow \Uparrow)$ and $\Delta \sigma_{\perp} = \frac{d^2\sigma}{d\Omega dE'} (\downarrow \Rightarrow - \uparrow \Rightarrow)$.  The two polarized cross section differences correspond to the target proton spin parallel and perpendicular to the incoming electron spin, respectively.   The slightly differing kinematics, influenced by the strong target magnetic field, did not permit the combination of data sets at the polarized cross section difference level for the setting where we have both longitudinal and transverse data, so the structure functions were formed using a model input according to:
\begin{align}
g_1(x,Q^2) & = K_1\bigg{[}\Delta \sigma_{\parallel}{\bigg(}1+\frac{1}{K_2}\mathrm{tan}\frac{\theta}{2}{\bigg)}{\bigg]}  + \frac{2g_2\mathrm{tan}\frac{\theta}{2}}{K_2y}\\
\label{g2fromg1}
g_2(x,Q^2) & = \frac{K_1y}{2}\bigg{[}\Delta \sigma_{\perp}{\bigg(}K_2+\mathrm{tan}\frac{\theta}{2}{\bigg)}{\bigg]} - \frac{g_1y}{2}\,,
\end{align}
where the kinematic terms, $K_1$ and $K_2$, are defined as
\begin{align}
K_1 & = \frac{MQ^2}{4\alpha}\frac{y}{(1-y)(2-y)}\\
K_2 & = \frac{1+(1-y)\mathrm{cos}\theta}{(1-y)\mathrm{sin}\theta}\,,
\end{align}
and $\theta$ is the angle of the scattered electron, $y = \nu/E$ and $\nu = E' - E$. A model~\cite{HallBRes} based on the CLAS Hall B data was used as the $g_1$ input for the extraction of $g_2$, except in the $Q^2$ = 0.05 GeV$^2$ setting where measured $\Delta\sigma_{\parallel}$ and $\Delta\sigma_{\perp}$ were used to solve the above for $g_1$ and $g_2$.  Details on the extraction of the polarized cross section differences can be found in the Methods section.

\begin{figure}[t!]
\includegraphics[angle=0,width=0.48\textwidth]{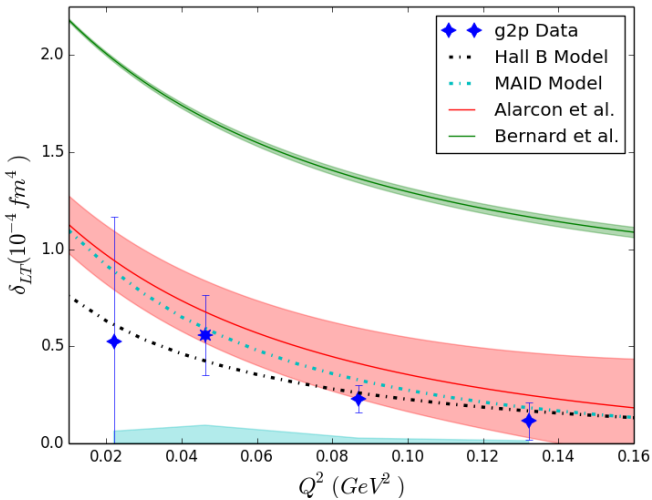}
\caption{\label{fig:dlt}The longitudinal-transverse spin polarizability for the proton as a function of $Q^2$, compared to existing world data~\cite{ELSA,EG1b}, phenomenological models~\cite{MAID2007,HallB} and $\chi$PT calculations~\cite{Gold,Krebs}. The $\delta_{LT}$ point indicated by an 8-pointed marker near $Q^2=0.05$ $GeV^2$ includes both $g_1$ and $g_2$ from E08-027 data, while the other three points use the CLAS model for the $g_1$ part of the integral. The cyan shaded region represents the systematic uncertainty.}
\end{figure}

\begin{figure}[t!]
\includegraphics[angle=0,width=0.48\textwidth]{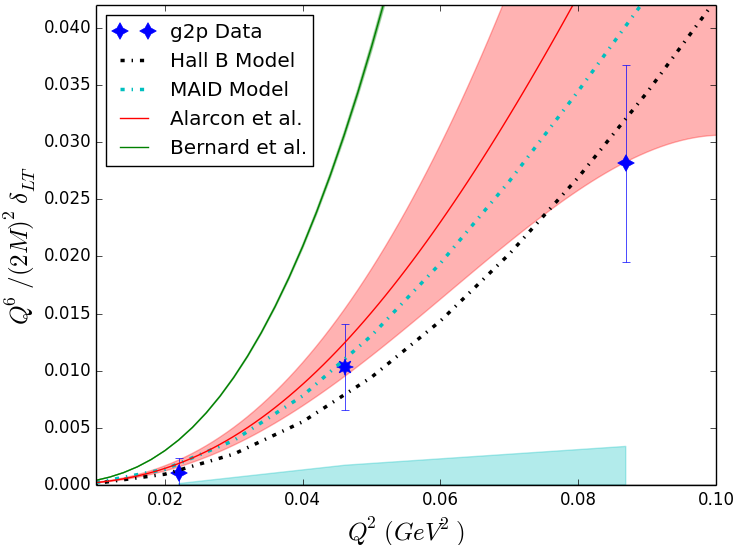}
\caption{\label{fig:dlt6}The longitudinal-transverse spin polarizability for the proton as a function of $Q^2$, compared to existing world data~\cite{ELSA,EG1b}, phenomenological models~\cite{MAID2007,HallB} and $\chi$PT calculations~\cite{Gold,Krebs}. The $\delta_{LT}$ point indicated by an 8-pointed marker near $Q^2=0.05$ $GeV^2$ includes both $g_1$ and $g_2$ from E08-027 data, while the other two points use the CLAS model for the $g_1$ part of the integral. The cyan shaded region represents the systematic uncertainty. On this plot the moment is scaled by $\frac{Q^6}{(2 M)^2}$ to form a unitless quantity, and is zoomed in on the lowest three $Q^2$ points.} 
\end{figure}
 
 The experimental cross section, calculated only for the longitudinal setting, was formed by normalizing the detected electron counts by target density and thickness ($\rho$), spectrometer acceptance ($V_{\mathrm{acc}}$), detector efficiencies ($\epsilon_{\mathrm{det}}$), livetime ($LT$) and accumulated charge ($Q/e$)  : \begin{eqnarray}
\sigma_0 = \frac{d^2\sigma}{d\Omega dE'}= 
\frac{N_{\mathrm{det}} }{Q/e \cdot \rho \cdot LT \cdot \epsilon_{\mathrm{det}} \cdot V_{\mathrm{acc}}}\,. 
\end{eqnarray}
 The spectrometer acceptance is defined with solid angle $\Omega$ and scattered electron energy $E'$ and was determined using a Monte-Carlo simulation~\cite{Chao3}. The same dilution factor in the asymmetry was applied to the cross section to obtain a pure proton result. Large systematics in the transverse cross sections made it preferable to form the polarized cross sections differences using the asymmetries from g2p data, and an unpolarized cross section from the Bosted-Christy model \cite{Bosted:2010}. The longitudinal cross section was used to determine how well the model agreed with the g2p data, and obtain an associated systematic error. It was determined from this comparison that the structure of the model matched our data very well, but  needed to be scaled by a factor of 
 $\approx 1.15$. 
 This scaling factor is perhaps not surprising due to the small amount of existing low $Q^2$ proton data available to constrain the model, and is in any case consistent within error bars with the E61 data~\cite{HallCRes} that was originally used to create the Bosted-Christy model. 
 This scaling factor is trusted to within the 9\% relative uncertainty of our measured cross section. An additional small uncertainty associated with structure differences between our data and the model brings the uncertainty of this method to around 10\%. However, the impact of this scaling factor on the higher moments is suppressed. We have  calculated it to be less than a 6\% relative uncertainty, which contributes to the total uncertainty at the same order as the dilution factor.
 
 
 The dilution factor uncertainty is a product of uncertainty propagated from the calculation of the packing fraction and uncertainty associated with acceptance effects; this systematic averaged between 6-8\%. The uncertainty in the reconstructed electron scattering angle is a product of the BPM and spectrometer optics uncertainties; this angle uncertainty is approximately 2\% and when propagated through the Mott cross section at forwards angles, gives a systematic uncertainty of 5\%. The beam and target polarizations contribute an uncertainty around 5\%, and the radiative corrections and elastic tail subtraction each contribute around 3\%. The uncertainty associated with the $g_1$/$g_2$ input to the polarized XS differences was 2\% or less. The contribution of the out of plane angle, the adjustment to constant $Q^2$, the charge normalization, and detector efficiencies were all found to contribute 1\% uncertainty or less to the total systematic. Adding the above errors in quadrature produces an approximately 14\% systematic uncertainty in each kinematic setting for the structure function, which is reduced slightly to 12\% in the moments by the kinematic weighting.

 Radiative corrections were applied using a combination of the Mo and Tsai (external corrections)~\cite{MT}  and POLRAD (internal corrections) formalisms~\cite{POLRAD,POLRAD4}. The polarized elastic tail contributions were calculated using the Ye and Arrington elastic form factor parametrization~\cite{Ye:2017gyb}. Inelastic corrections were calculated using an iterative, unfolding procedure. All unpolarized corrections were calculated under the energy peaking approximation~\cite{Stein}. A combination of g2p data and polarized model inputs (MAID and CLAS~\cite{MAID2007,HallB}) were incorporated into the process to improve the systematic uncertainty. Extrapolations in the iterative procedure were carried out at lines of constant $W$ where the input spectra exhibited the same angular dependence as the unfolded cross section~\cite{Zielinski:2017gwp}.
 
 Our results for the spin structure functions evaluated at a constant $Q^2$ are shown in Fig.~\ref{fig:SSF} as a function of invariant mass $W$. Blue stars are $g_2$ results from our perpendicular polarized cross section differences. For the moments shown in this paper, data which includes a model for the $g_1$ part shall be represented with these 4-pointed stars. Data which instead includes the E08-027 $g_1$ data (not displayed here) shall be represented with 8-pointed stars. In all figures, the error bars are the statistical uncertainty, and the blue shaded region beneath the data is the systematic uncertainty. Numerical values for the moments discussed in this article are found in the supplemental materials.

  The  adjustment to a constant momentum-transfer assumes the $Q^2$ dependence of the CLAS model~\cite{HallB}, although the MAID model predicted a similar evolution.  For all our spectra the correction was small compared to the SSF statistical uncertainty.  The method is described in detail in Ref~\cite{Zielinski:2017gwp} and contributed less than 1\% systematic uncertainty.
  
  These constant $Q^2$ structure functions were used to form the moments defined in Eqs.~\ref{DLTEQ} \&~\ref{D2EQ}.
   Our measured SSF data extends down to atleast $x$ = 0.05 for all $Q^2$.  This combined with the $x^2$-weighting of the integrals ensures 
   that the moments are largely insensitive to any unmeasured contribution at lower x.   
  
\begin{figure}[t!]
\includegraphics[angle=0,width=0.48\textwidth]{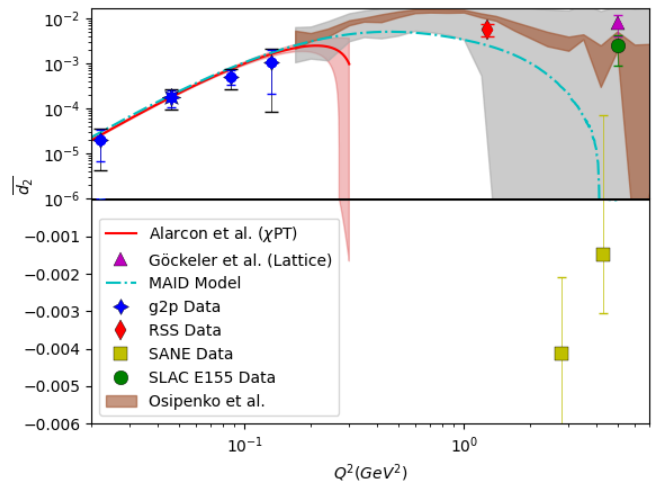}
\caption{\label{fig:d2}Higher moment $\overline{d_2}$ for the proton as a function of $Q^2$, compared to existing data~\cite{RSS,warmstrong,Anthony:2002hy}, the phenomenological models~\cite{MAID2007,Osipenko:2005nx}, $\chi$PT calculations ~\cite{Gold} and a Lattice QCD calculation~\cite{Gockeler}. The brown band is the Osipenko $et$ $al.$ model and the grey region is the associated error band. The $\overline{d_2}$ point indicated by an 8-pointed marker near $Q^2=0.05$ $GeV^2$ includes both $g_1$ and $g_2$ from E08-027 data, while the other three points use the CLAS model for the $g_1$ part of the integral. The region above zero is shown on a log scale on the y axis to clearly show the comparison with the model, while the negative half of the plot is shown on a linear scale to allow the proper inclusion of the SANE \cite{warmstrong} data.  The statistical (total) uncertainty of our meausurement is represented with the inner (outer) error bars.}
\end{figure}

Our results for the longitudinal-transverse spin polarizability, $\delta_{LT}$, in Fig.~\ref{fig:dlt}, represent the first ever experimental determination of this quantity for the proton. Fig.~\ref{fig:dlt6} is scaled from the same data to form a unitless quantity, which emphasizes the lowest $Q^2$ data as in Ref.~\cite{Drechsel2}. Comparisons were made to relativistic Baryon-$\chi$PT calculations from Alarcon $et$ $al.$~\cite{Gold} and Bernard $et$ $al.$~\cite{Krebs} and are represented by the red and green bands. The solid lines represent the central values while the shaded bands represent the calculation's uncertainty. Within our uncertainties, we show general agreement with the Alarcon $et$ $al.$ calculation and disagreement with the corresponding Bernard $et$ $al.$ calculation. 

One known difference between the two calculations rises from the inclusion of the $\Delta$(1232) resonance through a perturbative expansion. Bernard $et$ $al.$ adhere to an $\epsilon$ expansion scheme, wherein the mass difference between the $\Delta$-resonance and the proton ($\varDelta$) is assumed to be of similar scale to the pion mass $m_{\pi}$, allowing them to be expanded to the same order. Alarcon $et$ $al.$ use a $\delta$ power counting scheme, which assumes that the ratio of these parameters $\frac{m_{\pi}}{\varDelta}$ is similar to the ratio of $\varDelta$ and the nucleon mass $\frac{\varDelta}{M_p}$~\cite{DeltaCount}. It is interesting to note that the $Q^2$ dependence of the calculations are similar, but the normalizations appear to differ substantially as shown in Fig.~\ref{fig:dlt}. As Ref.~\cite{Gold} discusses, such a difference at $Q^2$ = 0 could result from their enforcement of ``consistent'' couplings to the Delta field, as opposed to Ref.~\cite{Krebs} where that consistency is not enforced. Additionally, the Bernard $et$ $al.$ results include only the leading-order predictions, and the recent analysis in Ref.~\cite{thurmann} indicates that higher order corrections are expected to be large. A comparison which accounts for these additional corrections is necessary to completely understand the underlying differences between the two theoretical approaches. 


Our results for the higher moment $\overline{d_2}$ are shown in Fig.~\ref{fig:d2}. These data show good agreement with the MAID model~\cite{MAID2007} and the calculation of Alarcon $et$ $al.$~\cite{Gold}. The brown (and grey) shaded regions are a result of using the method of Osipenko $et$ $al.$~\cite{Osipenko:2005nx} to generate a $\overline{d_2}$ result. The SANE~\cite{warmstrong} data at high $Q^2$ indicates the intriguing possibility of a zero crossing at large $Q^2$, while our data is consistent with the positive sign of the 
RSS~\cite{RSS} and SLAC~\cite{Anthony:2002hy} data, as well as a lattice QCD~\cite{Gockeler} calculation. On this plot, the region above zero is shown on a log scale to make the structure of our data visible, and the region below zero is shown on a linear scale to allow the inclusion of the SANE data. Our data shows the expected trend towards zero at the real photon point. Confirmation of the maximum predicted by the phenomenological models around 1 GeV$^2$ will require further measurements of $g_2$ in this region.

These polarizabilities provide new insight into the non-perturbative regime of the proton. The $\overline{d_2}$ results show good agreement with phenomenological models and the Alarcon~\cite{Gold} $\chi$PT calculation and suggest that a new medium $Q^2$ measurement is important to understand the transition region.  
Our results allow for unambiguous discrimination between two state-of-the-art $\chi$PT calculations for the longitudinal transverse spin polarizability, supporting the calculation of Alarcon $et$ $al.$~\cite{Gold}. In contrast to the original ``$\delta_{LT}$ puzzle'', the proton calculations show better agreement with the data than was originally observed for the neutron, although our data clearly favors one approach. This data represents a new benchmark for high precision discrimination of theoretical calculations in the strong QCD domain.


We would like to thank the Hall~A technical staff and
the accelerator operators for their efforts and dedication. We would also like to thank Jose Manuel Alarc\'{o}n, V\'{e}ronique Bernard, Peter Bosted, Eric Christy, Alexandre Deur, Evgeny Epelbaum, Franziska Hagelstein, Hermann Krebs, Sebastian Kuhn, Vadim Lensky, Ulf-G. Mei{\ss}ner, and Vladimir Pascalutsa for their very helpful discussion and suggestions on this publication.
This work was supported by the Department of Energy under grants DE-FG02-88ER40410, DE-FG02-96ER40950 and DE-AC02-06CH11357. The Southeastern Universities Research Association  operates the Thomas Jefferson National Accelerator Facility for the DOE under contract DE-AC05-06OR23177.
\newpage
\clearpage
\section{Methods}
The polarized continuous-wave (CW) electron beam was created by stimulating photoemission from a strained GaAs cathode using circularly polarized light. The polarization was flipped at 960 Hz in a pseudo-random fashion. The  helicity sequence was quartet pattern of ($+--+$) or ($-++-$) to minimize linear background effects~\cite{Chao3}. An insertable half-wave plate controlled the overall sign of the beam polarization and was flipped throughout the experiment to suppress helicity-dependent systematic effects. The beam polarization was measured using a M{\o}ller polarimeter. The average polarization was 84.0 $\pm$ 1.5\%~\cite{Moller,MollerRaw}.

The beam current was kept well below 100 nA, this current limit and a raster system designed to spread the beam spot out to a circle 2 cm in diameter were used to minimize target depolarization. New beam position monitors and read-out electronics were designed for the experiment to deal with the lower beam current, spiral raster pattern and chicane magnets installed to transport the beam through the large transverse target field~\cite{Musson}. This equipment achieved uncertainties of 1$-$2 mm and 1$-$2 mrad in the beam position and beam angle, respectively~\cite{Zhu}.

Polarized protons were created by means of the dynamic nuclear polarization (DNP)~\cite{PolHyper} of solid ammonia (NH$_3$) target beads, kept at a temperature of about 1 K in a $^4$He evaporation refrigerator. In DNP, the polarization enhancement is achieved via microwave stimulated transitions. Polarization is measured by a nuclear magnetic resonance (NMR) system, a high precision resistance-inductance-capacitance (RLC) circuit capable of detecting photons emitted or absorbed by proton spin flips as a change in the circuit's inductance.  A superconducting magnet run at 5 T and 2.5 T provided the necessary field strengths for the DNP process and the magnet system was rotatable for parallel and transverse polarization states. Average target polarizations were 70\% and 15\% for the 5 T and 2.5 T configurations, respectively~\cite{TargetPol}.

The scattered electrons were detected by the Hall A high resolution spectrometers~\cite{CEBAF}. A room temperature septum magnet was placed in front of the entrance to each spectrometer and decreased the minimum accepted scattering angle from 12.5$^{\circ}$ to 6$^{\circ}$.  Drift chambers tracked the electron trajectories and a pair of segmented plastic scintillators formed the data acquisition trigger. Particle identification was provided by a gas \v Cerenkov detector and a two-layer electromagnetic calorimeter. Efficient organization of readout electronics reduced the processing time and increased the achievable trigger rate to 6 kHz with $<$20\% deadtime. 

The polarized cross section differences used to form the structure functions were calculated from the product of experimental asymmetries ($A^{\mathrm{exp}}_{\parallel, \perp}$) and unpolarized cross sections ($\sigma_0^{\mathrm{exp}}$)
\begin{align}
\label{eq:poldiff}
\Delta\sigma^{\mathrm{physics}}_{\parallel, \perp} &= \Delta\sigma^{\mathrm{exp}}_{\parallel, \perp}  - \Delta\sigma^{\mathrm{tail}}_{\parallel, \perp} + \delta(\Delta\sigma^{\mathrm{RC}}_{\parallel, \perp})
\end{align}
where $\Delta\sigma^{\mathrm{exp}}_{\parallel, \perp} = 2\sigma_0A_{\parallel, \perp}$ and the radiative correction terms  $ \Delta\sigma^{\mathrm{tail}}_{\parallel, \perp}$ and $\delta(\Delta\sigma^{\mathrm{RC}}_{\parallel, \perp})$ represent the polarized elastic tail subtraction and polarized inelastic radiative corrections. This physics polarized cross section difference can be approximated as the Born cross section, as the higher order effects not included in the radiative corrections are heavily suppressed.

The asymmetries were formed according to
\begin{equation}
 A^{\mathrm{exp}} = \frac{1}{f \cdot P_t \cdot P_b}{\bigg (}\frac {Y_+ - Y_-}{Y_++Y_-}{\bigg )}
 \end{equation}
 with $Y_{\pm}= \frac{N_{\pm}}{LT_{\pm}Q_{\pm}}$ as the livetime (LT) and charge (Q) corrected counts (N) for each electron helicity state. Target and beam polarization is accounted for in $P_t$ and $P_b$, respectively. The dilution factor, $f$, corrects for contributions from unpolarized background. It was determined from the ratio of experimental scattering data collected on the ammonia target cell, empty target cell, liquid $^4$He, and a thin $^{12}$C disk. The carbon data were used to check and scale the Bosted-Fersch empirical fit~\cite{N2Scale}. This fit was then used to model the nitrogen background contribution. For this experiment, the dilution factor was usually around 0.15, with some structure varying along W.
 
 At the transverse kinematic settings, an additional correction factor was applied to the asymmetries to account for the out-of-plane angle between the polarization and scattering planes. The correction was applied as 1/cos($\theta_{\mathrm{OoP}}$). The combination of the chicane magnets and target magnetic field caused the asymmetries to acquire a significant out-of-plane term. Determined from the BPM and scattered electron reconstruction, the angle ranged from $\theta_{\mathrm{OoP}}$ = 25$^{\circ}$$-$ 65$^{\circ}$  (5 T) and $\theta_{\mathrm{OoP}}$= 1$^{\circ}$ $-$ 20$^{\circ}$ (2.5 T). 
\newpage
\clearpage

\newpage

\bibliography{g2p_transverse_paper.bib}
\end{document}